# RESEARCH PAPER

[1]Aakash Garg (aakash1913193@akgec.ac.in) , [1]Anant Patel (anant@1913124@akgec.ac.in) ,
[1]Ankit Tyagi (ankit1912070@akgec.ac.in) ,[1]Divyansh Raj (divyansh1912012@akgec.ac.in),
[2]Mr.Gaurav Chaudhary (  )

[1]*UG Final Year Students, Department of Information Technology, AKGEC College, UP, India*
[2] *Associate Professor, Department of Information Technology, AKGEC College, U.P., India*

## Abstract

Blockchain, the backbone of Bitcoin, has recently gained a lot of attention. Blockchain functions as an immutable record that enables decentralized transactions.

Blockchain-based applications are sprouting up in a variety of industries, including financial services, reputation systems, and the Internet of Things (IoT), among others. However, many hurdles of blockchain technology, including scalability and security issues, have to be overcome.

Many industries, including finance, medicine, manufacturing, and education, use blockchain applications to capitalize on this technology's unique set of properties.

Blockchain technology (BT) has the potential to improve trustworthiness, collaboration, organization, identity, credibility, and transparency. We provide an overview of blockchain architecture, various different kinds of blockchain as well as information about the Decentralized apps which are also known as Dapps. This paper provides an in-depth look at blockchain technology.

## 1. Introduction

The concepts of bitcoin and blockchain were initially suggested in 2008 by someone using the pseudonym Satoshi Nakamoto, who detailed how cryptology and an open distributed ledger may be merged into a digital currency application (Nakamoto 2008).

Initially, bitcoin's unusually high volatility and many governments' attitudes about its complexity slowed its expansion slightly, but the benefits of blockchain—the underlying technology of bitcoin—attracted increasing interest.

Some of the benefits of blockchain include its distributed ledger, decentralization, information transparency, tamper-proof architecture, and openness.

Blockchain technology has been used in a variety of applications, including digital currency and finance, as well as health care, supply chain management, market monitoring, smart energy, and copyright protection.

Nowadays cryptocurrency has become a buzzword in both industry and academia. As one of the most successful cryptocurrencies, Bitcoin has enjoyed huge success with its capital market reaching 10 billion dollars in 2016 [1]. With a specially designed data storage structure, transactions in the Bitcoin network could happen without any third party [2].

Blockchain may be utilised in a variety of financial services, including digital assets, remittance, and online payment, since it allows payments to be completed without the involvement of a bank or an intermediary [3], [4].

It may also be used in other fields such as smart contracts [5], public services [6], and the Internet of Things (IoT) [7].The financial industry appears to have played a pioneering role because cryptocurrencies were the first practical blockchain applications.

Nonetheless, the promise of this technology has piqued the interest of other fields in recent years, resulting in a slew of new projects. BT is still in its early stages, with few generally established standards and frameworks.

The rest of this paper is organized as follows. Section II introduces blockchain technology. Section III shows typical consensus algorithms used in blockchain. Section IV summarizes the uses of Dapps and kinds of blockchains. Section V discusses some possible future directions and section VI concludes the paper

## 2. Blockchain Technology

In this section, we will present essential understanding regarding blockchain technology (BT) by quickly describing it, its properties, and functionalities (section 2.1)

### 2.1. Overview

When discussing BT, distributed ledger technology must be stated because it is an umbrella word that encompasses blockchains as one form (Benčić and Podnar Žarko, 2018).

A distributed ledger employs separate systems (nodes) to record, share, and coordinate transactions in a decentralised network (Kakavand et al., 2017).A blockchain works similarly, but it organises its data into blocks that are cryptographically and chronologically linked together, and it may also employ various types of consensus processes and smart contracts (Anwar, 2019).

Haber and Stornetta described a cryptographically protected chain for the BT in 1991 (Haber and Stornetta, 1991), then in 1993 they and others expanded that notion with specific capabilities like as timestamping (Bayer et al., 1993). Their architecture still contained issues, such as the double-spending issue (Chohan, 2018) and the requirement for a trusted third to validate all transactions.

Under the alias "Satoshi Nakamoto," a whitepaper describing the innovative peer-to-peer digital currency "Bitcoin" that fixed these problems was published in 2008.The Bitcoin network finally went online in 2009 and saw a wild ride in terms of market value (shortly over 20,000$) [8] and media attention.

It received the most notoriety as a result of the numerous news stories regarding the growth of its worth. Since 2009, a large number of cryptocurrencies have been created (there are currently over 2,000 distinct types), and BT has come to be recognised as a technology that not only can offer an infrastructure to handle currencies but also is allowing the implementation of a large number of use cases.

## 2.2. Blockchain Architecture

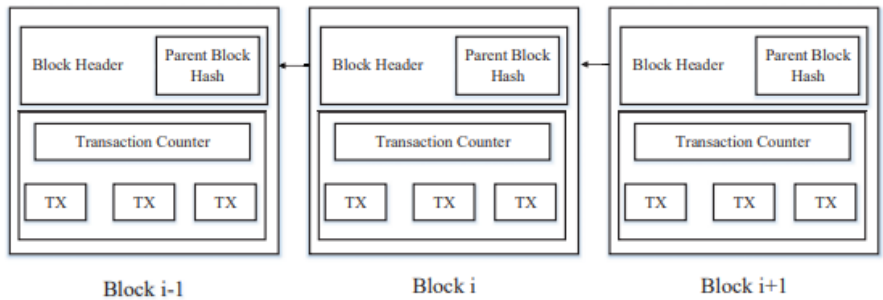

Fig. 1: An example of a blockchain which consists of a continuous sequence of blocks.

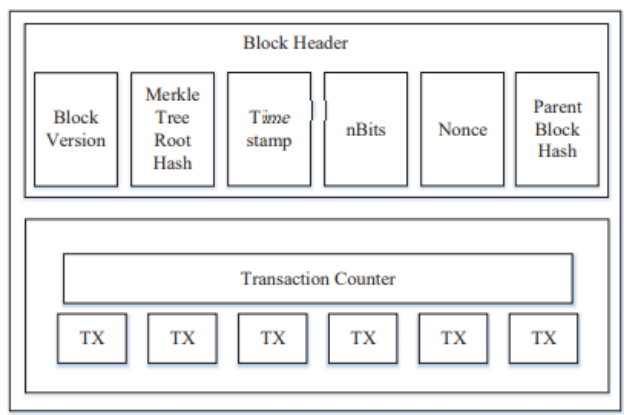

Fig. 2: Block structure

Blockchain is a series of blocks that, like a traditional public ledger, contains an exhaustive list of transaction records [9]. A blockchain is shown as an example in Figure 1.

A block only has one parent block if the block header contains a preceding block hash. It is important to note that uncle blocks' hashes—those of the block's forebears—would likewise be kept on the Ethereum network [10].

The genesis block, which has no parents, is the very first block on a blockchain.

A. *Block*

As shown in Figure 2, a block comprises of the block header and the block body. The block header contains the following information:

(i) Block version: identifies the set of block validation rules to follow.
(ii) Merkle tree root hash: the sum of all the block's transactions.
(iii) Timestamp: the time right now expressed in seconds of universal time since January 1, 1970.

- (iv) nBits: target limit for a legitimate block hash.
- (v) Nonce: a 4-byte field that typically starts at 0 and grows with each hash computation.
- (vi) Parent block hash: A 256-bit hash value that refers to the prior block

A transaction counter and transactions make up the block body. Depending on the block size and the size of each transaction, a block can contain a maximum number of transactions. Blockchain employs asymmetric cryptography-based digital signatures in an unreliable setting.[11].

B. *Key Characteristics of Blockchain*

In summary, blockchain has following key characteristics.

• Decentralization

Conventional centralised transaction systems require that each transaction be verified by a single trusted third party (such as the central bank), which invariably leads to cost and performance bottlenecks at the central servers.In contrast to the centralised form, blockchain eliminates the necessity for third parties.Blockchain uses consensus techniques to preserve data consistency across distributed networks.

• Resilience.

Transactions can be verified fast, and sincere miners would not accept any invalid transactions. Once a transaction is added to the blockchain, it is very difficult to remove it or roll it back. Invalid transaction-containing blocks may be found right away.

# 3. Types of Blockchain Network

## 3.1. Public Blockchain Network

Public blockchain network is a permissionless network in which anyone can join or participate the network. In Public blockchain network, all users or nodes of the network have equal rights to access the network, create new blocks of data, and validate blocks of data.

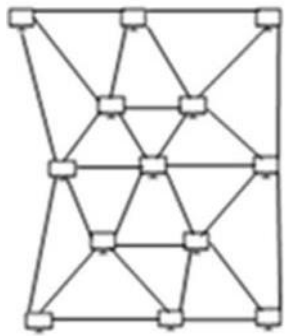

Fig.3: Public Blockchain Network

## 3.2. Private Blockchain Network

A private blockchain is a permissioned blockchain network in which only a single organization has access and authority over the network. In Private Blockchain, anyone who wants to join it must seek for the permission from the governing body of the blockchain, and the public people cannot pierce it.

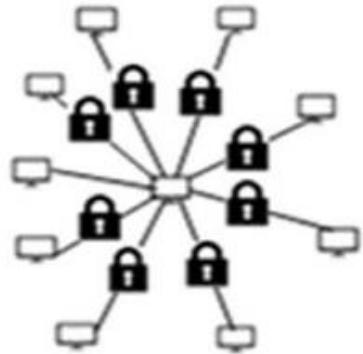

Fig.4: Private Blockchain Network

## 3.3. Consortium Blockchain Network

Consortium blockchain network are permissioned blockchain administrated by a group of organizations, rather than single entity, as in the case of the private blockchain network. Consortium blockchain network are more decentralized than private blockchain network therefore resulting in higher levels of security in case of consortium blockchain network.

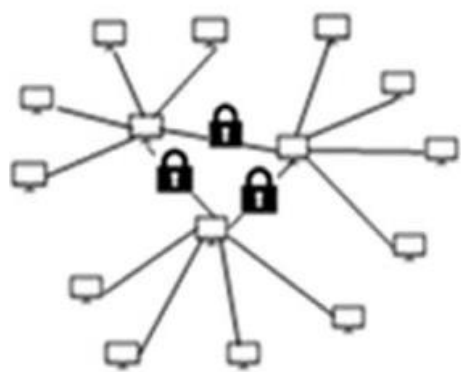

Fig.5: Consortium Blockchain Network

# 4. dApps

The software programs using a blockchain are called "decentralized applications" or "dApps", and are one of the main new trends in software development. A search of scientific and technical documents made with Google Scholar in July 2021 found 36,700 results for "smart contracts" development, a number higher or much higher than the results for microservices development (20,500), global software engineering (7670), devops development (23,500), and even IoT "software development" (30,400).

Decentralized applications (dApps) are digital applications or programs that exist and run on a blockchain or peer-to-peer (P2P) network of computers rather than on a single computer. DApps (also known as "dApps") are outside the scope of influence and control of a single authority. DApps, often built on the Ethereum platform, can be built for a variety of purposes, including gaming, finance, and social media.[13]

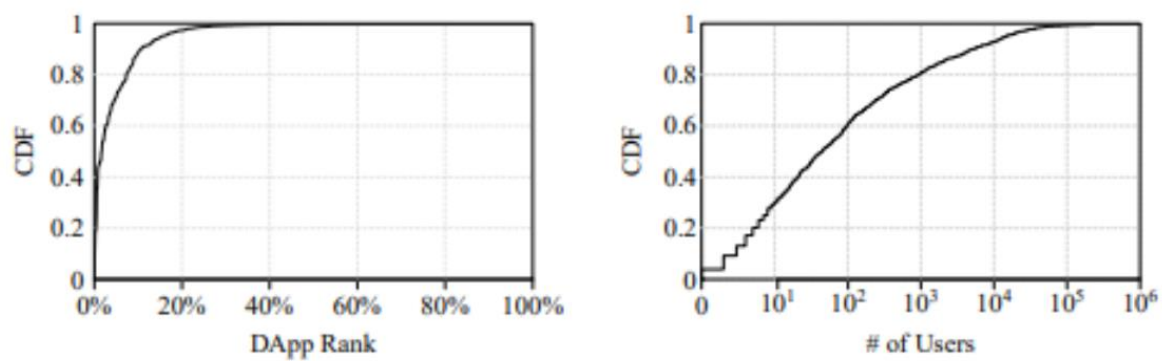

Fig.6: dApps popularity by users[19]

## 4.1 Classification of dApps

There are several characteristics that can classify decentralized applications. This whitepaper categorizes DApps based on whether they have their own blockchain or use another DApp's blockchain. Based on this criteria, there are three types of dapps.[14]

Type I decentralized applications have their own blockchain. Bitcoin is the most prominent example of a Type I decentralized application, but Litecoin and other "altcoins" are of the same type.

Type II Decentralized Application uses the Blockchain of Type I Decentralized Application. These applications have protocols and tokens which are important for function.

Type III distributed applications use the protocol of distributed type II applications. Type III distributed applications are protocols and have the tokens they need to function. For example, the SAFE network, which uses the Omni protocol to issue "safecoins" that can be used to purchase decentralized file storage, is an example of a Type III decentralized application.

| Category | DApps # | DApps % | Users # | Users % | Transactions # | Transactions % | Transaction Volume # | Transaction Volume % |
|---|---|---|---|---|---|---|---|---|
| Exchanges | 71 | 7.1% | 778,031 | 35.4% | 13,708,713 | 45.9% | 5,570,026.10 | 61.5% |
| Games | 294 | 29.5% | 184,730 | 8.4% | 5,834,574 | 19.5% | 226,506.11 | 2.5% |
| Finance | 93 | 9.3% | 517,525 | 23.5% | 2,571,729 | 8.6% | 2,321,199.88 | 25.6% |
| Gambling | 154 | 15.5% | 77,790 | 3.5% | 1,781,856 | 6.0% | 448,476.51 | 5.0% |
| Development | 30 | 3.0% | 269,821 | 12.3% | 1,154,346 | 3.9% | 20,525.36 | 0.2% |
| Storage | 13 | 1.3% | 249,544 | 11.3% | 1,031,779 | 3.5% | 11.29 | 0.0% |
| High-risk | 130 | 13.1% | 47,186 | 2.1% | 965,131 | 3.2% | 370,543.94 | 4.1% |
| Wallet | 17 | 1.7% | 193,790 | 8.8% | 787,160 | 2.6% | 2,020.50 | 0.0% |
| Governance | 18 | 1.8% | 188,201 | 8.6% | 633,211 | 2.1% | 132.29 | 0.0% |
| Property | 24 | 2.4% | 46,707 | 2.1% | 485,341 | 1.6% | 47,421.00 | 0.5% |
| Identity | 12 | 1.2% | 82,251 | 3.7% | 425,860 | 1.4% | 5,330.45 | 0.1% |
| Media | 48 | 4.8% | 127,315 | 5.8% | 403,055 | 1.4% | 1,144.43 | 0.0% |
| Social | 72 | 7.2% | 88,355 | 4.0% | 381,534 | 1.3% | 10,065.19 | 0.1% |
| Security | 14 | 1.4% | 40,684 | 1.9% | 127,550 | 0.4% | 17,211.19 | 0.2% |
| Energy | 3 | 0.3% | 21,312 | 1.0% | 95,025 | 0.3% | 22,127.17 | 0.2% |
| Insurance | 1 | 0.1% | 5,755 | 0.3% | 19,575 | 0.1% | 0.52 | 0.0% |
| Health | 2 | 0.2% | 4 | 0.0% | 9 | 0.0% | 0.00 | 0.0% |
| All DApps | 995 | | 2,199,059 | | 29,846,075 | | 9,057,344.36 | |

Table I: dApps popularity by categories (Categories are sorted by transactions) [19]

## 4.2 Frameworks of dApps

### 4.2.1 Ripple

This platform will be used to develop decentralized blockchain applications.x

It also helps facilitate quick and cheap transactions. Despite being a cryptocurrency, people can develop their currency easily by using the network of RippleNet. This framework enables users to find new users in new markets, expand their services, and provide a superior user experience worldwide. It makes transactions faster and easier, especially when it comes to cross-border payments.[17]

### 4.2.2 Hedera Graph

This allows developers to create a whole new class of dApps with scalability and the ability to manage thousands of transactions. It provides the highest level of security and helps build fair, fast and secure apps, whether you are an enterprise or a start-up, and even go beyond blockchain to create decentralized apps. It uses asynchronous Byzantine fault tolerance to act as a consensus mechanism.[17]

### 4.2.3 Hardhat

Hardhat's intelligent contract development environment gives developers the right tools to manage their development workflow. Furthermore, Hardhat also ensures efficiency by introducing automation at certain steps along with the possibility of new productive functions. Hardhat comes with a pre-built local Ethereum network aligned with the core goals of development. The hard hat and robustness equation is evident in the framework's focus on debugging robustness.[18]

## 5. Choosing the blockchain platform

We define as dApp a software system that uses DLT, typically a blockchain, as a central hub to store and exchange information through SCs. A dApp is composed of SCs running on a blockchain and of applications able to create and send transactions to them. These applications typically provide a human interaction interface, running on a PC or on a mobile device.

Identifying available blockchain platforms is the first step in choosing a blockchain platform for developing a business solution. There is currently no single source identifying a list of active blockchain projects. References to research publications were also not considered a successful approach. Searching by Internet search engines is the simplest and most straightforward process. Instead of an open web search, we recommend searching for technical articles published by domain-specific information technology publishers such as G2, Hacker Noon, DZone, ValueCoders, Gartner, Medium, LeewayHertz, ReadWrite, TechnoDuet. These articles are subject to change and new versions are often released. Therefore, visit technical articles from these publishers for the latest list of blockchain platforms. However, some blockchain projects are still in the proposal or incubation stage, and some are retired or in the lifecycle stage, so it's a good idea to clean up the list before evaluating. To do. It is often difficult to determine the status of a project solely from technical articles published by third parties. Therefore, we recommend that you visit her website for each blockchain project and select only active projects. If the current project version has a Long Term Support (LTS) release, that's an added plus.[15]

Different blockchain framework are Ethereum, Hyperledger, EOS, R3Corda, Quorum.[16]

## 6. Conclusions and future work

The key reason to use a blockchain is trust. If a system can be developed and deployed by an organization and its users trust this organization, there is no reason to use a blockchain.

In the case that it is not possible to trust a single organization managing the system, which should be open to all participants—some of whom might try to attack or exploit the system—a public blockchain is the choice. In managing digital currencies and tokens, public blockchains like Bitcoin, Ethereum, and many others have proved to be very effective and reliable.